\documentclass[aps,prl,twocolumn,superscriptaddress,floatfix,longbibliography]{revtex4}
\usepackage{amsfonts}
\usepackage{amssymb}
\usepackage{graphicx}
\usepackage{dcolumn}
\usepackage{bm}
\usepackage{amsmath}
\usepackage[colorlinks,linkcolor=magenta,citecolor=blue,urlcolor=blue]{hyperref}
\usepackage{changes}
\usepackage{float}

\begin{document}

\title{Scale-free localization versus Anderson localization in unidirectional quasiperiodic lattices}
\author{Yu Zhang}
\affiliation{Beijing National Laboratory for Condensed Matter Physics, Institute
of Physics, Chinese Academy of Sciences, Beijing 100190, China}
\affiliation{School of Physical Sciences, University of Chinese Academy of Sciences,
Beijing, 100049, China}

\author{Luhong Su}
\affiliation{Department of Physics, The Hong Kong University of Science and Technology, Clear Water Bay, Kowloon, Hong Kong, China}

\author{Shu Chen}
\email{schen@iphy.ac.cn}
\affiliation{Beijing National Laboratory for Condensed Matter Physics, Institute of
Physics, Chinese Academy of Sciences, Beijing 100190, China}
\affiliation{School of Physical Sciences, University of Chinese Academy of Sciences,
Beijing, 100049, China}
\date{\today }

\begin{abstract}

Scale-free localization emerging in non-Hermitian physics has recently garnered significant attention. In this
work, we explore the interplay between scale-free localization and Anderson localization by investigating a
unidirectional quasiperiodic model with generalized boundary conditions. We derive analytical expressions of the
Lyapunov exponent from the bulk equations. Together with the boundary equation, we can determine properties
of eigenstates and spectra and establish their exact relationships with the quasiperiodic potential strength and the
boundary parameter. While eigenstates exhibit scale-free localization in the weak-disorder regime, they become
localized in the strong-disorder regime. The scale-free and Anderson-localized states satisfy the boundary
equation in distinct ways, leading to different localization properties and scaling behaviors. Generalizing our
framework, we design a model with exact energy edges separating the scale-free and Anderson-localized states
via the mosaic modulation of quasiperiodic potentials. Our models can be realized experimentally in electric
circuits

\end{abstract}

\maketitle




{\color{blue}\textit{Introduction.}}
Non-Hermitian systems can exhibit some
novel localization phenomena without Hermitian counter-
parts. 
A paradigmatic example is the non-Hermitian skin
effect (NHSE), which manifests as the localization of majority of eigenstates at the boundary under open boundary
conditions (OBCs) \cite{Lee,Kunst,NHSE_Edge,Torres,Murakami}.
The NHSE is intrinsically related to the non-Hermitian point gap
 \cite{NHSE_Sato,NHSE_CF,Slager} and is highly sensitive to boundary conditions \cite{LeeCH2019,GuoCX2021}.
The interplay between
non-Hermiticity and disorder has created extreme richness
to localization phenomena in generic non-Hermitian systems\cite{Shindou2021,Ryu2021,JPA2009} and stimulated intensive studies on non-Hermitian systems with random disorder \cite{HNPRL,HNPRB,Nelson,AZ,Efetov,Ramezani,Hughes,HuangY2020,Ryu2021} or quasiperiodic disorder \cite{Avila2015,ZhouQ2023,jazaeri,longhiPRL,jiang2019interplay,longhi2019metal,LZC,zeng2019topological,Liu2020,Liutong,Liu2020L,Zeng2020,Cai2021AA,Longhi2021AA,Zhai2020,Liu2020DM,Cai2021PW,XuZH,DWZhang, Pxue,LWZhou,LWZhou1,XLGao,TongLiu,LiZ2024,WangL2024,Datta,Gandhi,Mishra,Halder}.
. Competition between the NHSE and Anderson
localization (AL) has been studied in Refs.\cite{jiang2019interplay,Hughes,LZC}, unveiling that the nonreciprocal hopping shifts the localization-delocalization transition point and induces a topological transition, where the rescaled transition point can be validated by the winding number \cite{longhiPRL,jiang2019interplay}.

Recently, a distinctive type of non-Hermitian localization,
dubbed as scale-free localization (SFL), was proposed \cite{SFL_Li} and has attracted intensive studies \cite{Li-NC,SFL_criticalNHSE,SFL_Guo,SFL_Wang,SFL_cpb,SFL_disorder,ZhangYi,SFL-exp,SFL-LiZ,SFL-LiFX,SFL-WZ,Yuce2024,GuoCX2024,Fulga,WangHF}. SFL is defined as
a special localization phenomenon in non-Hermitian systems,
characterized by a localization length which is proportional
to the system size. This feature stands in stark contrast to
the NHSE, where the localization length remains independent
of the system size. In SFL, however, the localization length
increases linearly with the system size. It is striking that even a
single non-Hermitian impurity in a Hermitian bulk lattice can
generate a significant number of SFL modes \cite{SFL_Guo},  which challenges our conventional understanding of localization. 
The simplest model realizing SFL is the unidirectional hopping model \cite{SFL_cpb,SFL_disorder}, which describes a system with hopping in only one direction. This model  can be solved analytically \cite{GuoCX2021} and is crucial for understanding the origin of SFL.

In this work, we investigate the interplay of SFL and
AL by considering the unidirectional hopping model with
various quasiperiodical potentials. In unidirectional hopping
lattices, due to the absence of reflection, one may wonder
whether Anderson-localized states can emerge, as usually the
Anderson localization phenomenon arises from a destructive
interference effect\cite{AL}. . Although a recent work numerically
demonstrated the existence of localized states in the unidirec-
tional hopping lattice with random disorder  \cite{SFL_disorder}, a theoretical understanding on the mechanism for the formation of localization and the transition between SFL and localized states is
still lacking.

\begin{figure}
\centering
\includegraphics[width=1\linewidth]{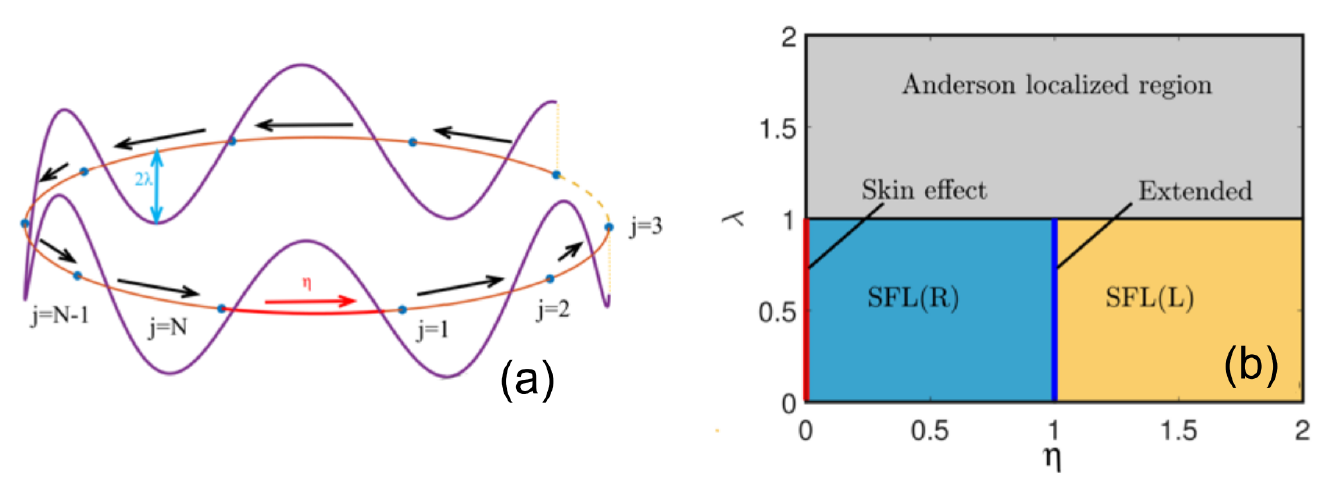}
\caption{
 (a) Schematic diagram of the unidirectional quasiperiodic model. (b) Schematic phase diagram of the unidirectional
quasiperiodic model with different strengths of the quasiperiodic
disorder  ($\lambda$) and boundary hopping term ($\eta$).
\label{Phasediagram}}
\end{figure}

Without loss of generality, we first consider the uni-
directional quasiperiodic model as schematically shown in Fig.\ref{Phasediagram}(a),  and demonstrate that the extension of the notation of the Lyapunov exponent (LE) can describe either the localized state or the SFL state. In the large-size limit, we get analytical expressions for the LE. Combining with the boundary
condition, we can determine properties of both eigenstates
and spectra, which allow us to study analytically the interplay
between SFL and AL. A schematic phase diagram is provided
in Fig.\ref{Phasediagram}(b). . The scaling behavior of eigenvalues and eigenstates varies across different regions. In the weak-disorder region, SFL persists but its properties are affected by the disorder strength and boundary conditions. The SFL can be
left-localized or right-localized depending on the value of the
boundary parameter. In contrast, the LE of the localized state
is determined by the bulk equation, whereas the boundary
equation gives the estimation of eigenvalues.
Furthermore, we study the unidirectional quasiperiodic model with mosaic modulation and unveil the existence of SFL-AL edges.

{\color{blue}\textit{Model and method.}}
We consider a unidirectional chain with quasiperiodic potentials described by the following Hamiltonian:
\begin{equation}
H=\sum_{j=1}^{L-1} t \left\vert
j+1\right\rangle \langle j|+\eta \left\vert 1\right\rangle \langle L|+ \sum_{j=1}^{L} 2 \lambda \cos ( 2 \pi \alpha j) |j\rangle \langle j|,
\label{Hamiltonian}
\end{equation}%
where $t$ represents the right hopping amplitude, $\alpha=\frac{\sqrt{5}-1}{2}$ is an irrational number,  and $\lambda$ quantifies the strength of the quasiperiodic potential.
The parameter $\eta$ is the strength of the hopping term connecting the first and the last sites.
Varying $\eta$ can interpolate the OBC ($\eta=0$)  to a periodic boundary condition (PBC) ($\eta=1$).  For arbitrary $\eta$, the boundary condition is referred to as a  generalized boundary condition (GBC). For convenience, we adopt $t = 1$ as the unit of energy and set $\eta \geq 0$ throughout the following discussions.

In the absence of disorder, i.e., $\lambda=0$, the model simplifies into the unidirectional hopping model with a GBC and can be solved analytically\cite{GuoCX2021,SFL_cpb,SFL_disorder}.
The eigenvalues satisfy $E^L-\eta=0$, resulting in an energy spectrum expressed as $$ E_m=\eta^{\frac{1}{L}} e^{i\theta_m}, (\theta_m=\frac{2m \pi}{L},m=1,2...L) .$$
The eigenstates can be written as
$$|\psi_m \rangle=c_0[1,\eta^{\frac{1}{L}} e^{i\theta_m},\eta^{\frac{2}{L}} e^{i 2\theta_m}\dots \eta ^{\frac{L-1}{L}} e^{i (L-1)\theta_m}]^T,$$
where $c_0$ is the normalization factor.
These eigenstates are localized at either the left or right edge with the localization length given by $\xi =L/|\ln(\eta)| $, which is proportional to the system size.
The model can be regarded as a minimal model to realize SFL eigenstates.

In the presence of the quasiperiodic potential, we can get
the eigenequation in the bulk from Eq.(\ref{Hamiltonian}), which reads as
\begin{equation}
E u_j = u_{j-1}+ 2\lambda \cos(2\pi \alpha j)u_j , \label{bulkeq}
\end{equation}
where $u_j$ is the eigenstate distribution at the $j$th site.
It gives the following recursion relation:
\begin{equation}
u_j=\frac{1}{E-2\lambda \cos(2 \pi \alpha j)} u_{j-1}, \label{eigenequation}
\end{equation}
in which $u_j$ only depends on $u_{j-1}$.
To study the localization property of the eigenstate, we introduce the LE ($\gamma$) defined by
\begin{equation}
|u_{j+l_d} |\approx  e^{- \gamma l_d} |u_{j}|,   \label{LE}
\end{equation}
where $l_d$ is the distance between different sites.
While $\gamma=0$  indicates a delocalized eigenstate,  $\gamma \neq 0$  indicates that the amplitude of  the eigenstate decays ($\gamma>0$) or grows ($\gamma<0$) with the increasing distance. The localization length $\xi$ is given by $\xi = 1/|\gamma|$.

Combining Eq.(\ref{eigenequation}) and (\ref{LE}), we get the expression for the LE:
$$
\gamma = \frac{1}{n} \sum_{j=1}^{n} \ln(|E-2\lambda \cos[2\pi \alpha (j+j_0)]|),
$$
where we assume the wave function starting from the ${j_{0}}$th site.
Because of the non-monotonic nature of the cosine function, the single wave function does not decay strictly exponentially over a short distance but exhibits exponential decay over a long distance on average.
If we consider a sufficiently long chain, we can make the transformation $2\pi \alpha (j+j_0) \rightarrow \theta$ and replace the sum with an integral \cite{Longhi2021AA,longhi2019metal}.
Then the LE can be expressed as
\begin{align}
\label{eq1} \gamma &= \frac{1}{2\pi}\int_{0}^{2 \pi} \ln(|E-2\lambda \cos(\theta)|)d\theta .
\end{align}

The eigenvalues of a non-Hermitian system can be complex.
Thus, we consider the LE in the entire complex plane:
\begin{equation}
\gamma =
\begin{cases}
  \ln(|\lambda|) & \text{If}  \quad |E| \leq  2\lambda,  \Im(E)=0,\\
  \ln |\frac{E+\sqrt{E^2-(2\lambda)^2}}{2}| & \text{otherwise}.
\end{cases} \label{LEE}
\end{equation}
It should be noted that this function is a multivalued function for complex $E$, and we have another restriction: $|\frac{E + \sqrt{ E^2-(2\lambda)^2 }}{2\lambda}|>1$,
which can also be written as
$
\gamma >\ln(|\lambda|).
$
The details of the calculation can be found in Supplement Material \cite{SM}.

Further, considering the GBC, we obtain the boundary
equation which connects the eigenstate distribution at the first
site to the last site:
\begin{align*}
 u_1= \frac{\eta}{E-2\lambda \cos(2\pi \alpha 1)}u_L . \label{bc}
\end{align*}
Using the recursion relation Eq.(\ref{eigenequation}), iteratively $L-1$ times, we get a self-consistent equation:
\begin{equation}
\eta \prod_{j=1}^L  \frac{1}{E-2\lambda \cos(2\pi \alpha j)} =1. \label{bdE}
\end{equation}
Using Eqs.(\ref{bdE}) and (\ref{LEE}), we can determine both the distribution of eigenvalues and the properties of the eigenstates. We shall show that there exist two different mechanisms to fulfill Eqs.(\ref{bdE}) and (\ref{LEE}) simultaneously in the SFL and AL regimes, respectively.

\begin{figure*}
    \centering
    \includegraphics[scale=0.65 ]{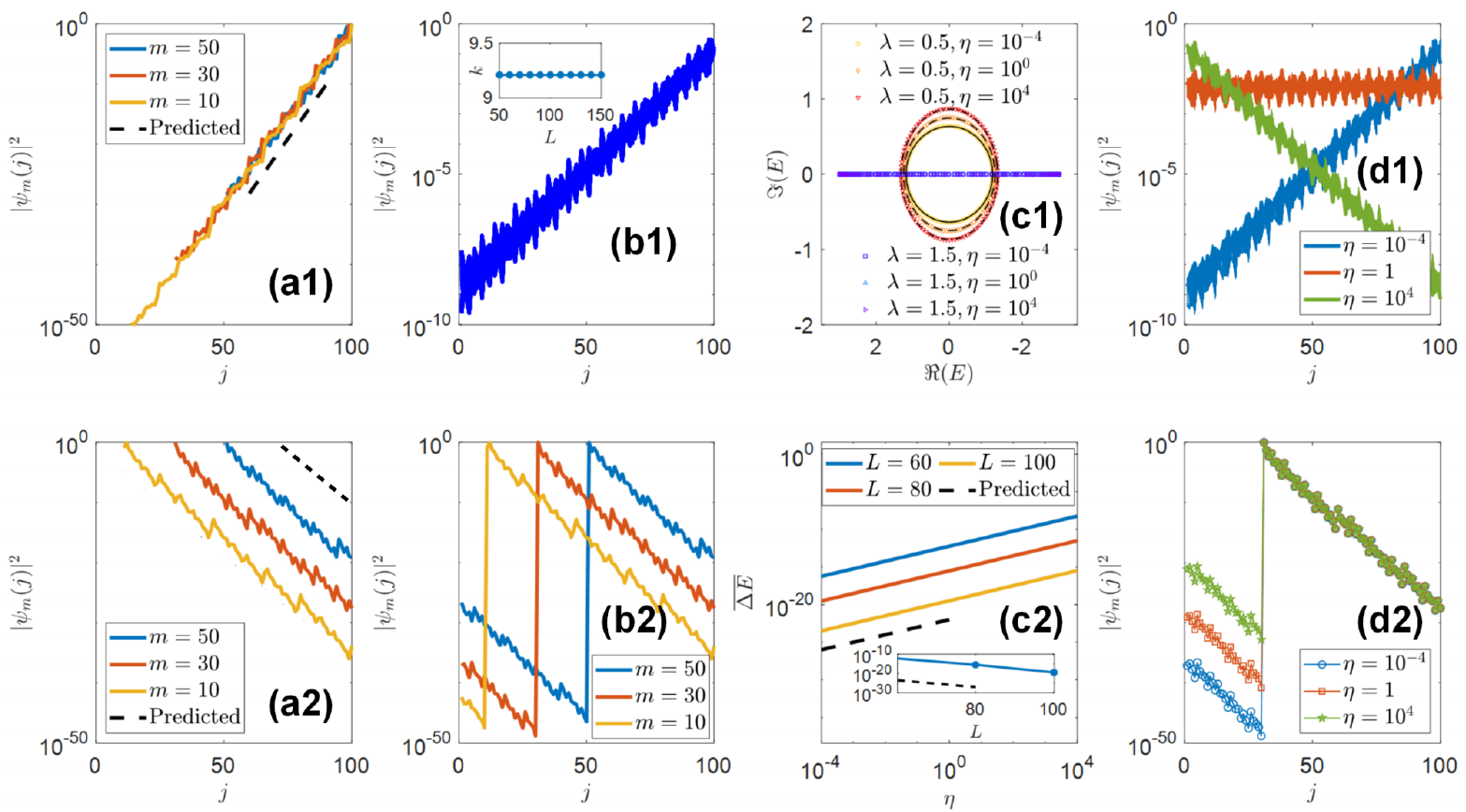}
    \caption{Eigenvalues and eigenstates for the unidirectional quasiperiodic model with different boundary conditions.
    [(a1),(a2)] Typical eigenstates for different disorder strengths under OBCs: $\lambda=0.5$ (a1) and $\lambda=1.5$ (a2).
The dashed line represents the decay of the wave function predicted by theoretical analysis.
    [(b1),(b2)] Eigenstates for systems under GBCs with $\eta=10^{-4}$: all eigenstates with $\lambda=0.5$ (b1) and typical eigenstates with $\lambda=1.5$ (b2).
    We  apply a linear fitting procedure to  $\ln(|\psi|)$ and $j/L$:
    $\ln(|\psi|)=k (j-j_0) /L+ c$.
     The inset in panel (b1) shows how the slope varies with system size.
     (c1) Spectra for the system with various $\lambda$ and $\eta$. The points are the eigenvalues from  exact
    diagonalization (ED)  and the black lines are the analytical results. The black solid line, the dashed line, and the dash-dotted line correspond to $\eta = 10^{-4}$, $\eta = 1$, and $\eta = 10^{4}$, respectively.
     (c2)  $\overline{\Delta E}$ versus $\eta$ for the system with $\lambda=1.5$ and different $L$. The inset shows $\overline{\Delta E}$ versus $L$. Both dashed black lines have the slopes predicted by our analytical results and are consistent with the results form ED.
     (d1) All eigenstates in the region of $\lambda=0.5$ with different boundary parameters. We choose $L=100$.
     (d2) Some typical eigenstates in the region of $\lambda=1.5$ with different boundary parameters. We choose $m=30$ and $L=100$. }
    \label{uniAA}
\end{figure*}

{\color{blue}\textit{Interplay  of SFL and localization.}}
We first consider the case under the OBC, which gives the boundary equation $E u_1 = 2\lambda \cos(2\pi \alpha )u_1$. It is shown that $E=2\lambda \cos(2\pi \alpha)$ is an eigenvalue of the system with the wave function of the eigenstate determined by using Eq.(\ref{eigenequation}) iteratively. 
From Eq.(\ref{bulkeq}), we see
$$
E_m= 2\lambda \cos(2\pi \alpha m),\ m=1\dots L
$$
are eigenvalues of the system.
The eigenstate corresponding to $E_m$ is confined to the region $[m,L]$ with $u_{i}=0$ for all $i< m$ and other $u_j$ with $j=m,\dots,L$ determined by using Eq.(\ref{eigenequation}) iteratively.
In Fig.\ref{uniAA} (a1) and \ref{uniAA}(a2), we display distributions of wave functions for several typical eigenstates in the regions of  $|\lambda|<1$ and $|\lambda|>1$, respectively.
For $|\lambda|<1$, all the eigenstates accumulate at the right boundary and terminate at the site $m$ with the LE  $\gamma<0$, indicating that these eigenstates are skin states.
For $|\lambda|>1$, the distribution of eigenstate decays exponentially in a unidirectional way from the $m$th site towards the right boundary.
These states are localized states with localized centers at the $m$th site.
Localization properties of both skin states and localized eigenstates can be characterized by the LE given by $\gamma=\ln(|\lambda|)$.
A transition from the NHSE to AL occurs at $|\lambda|=1$.
Although  the eigenvalues are given by the uniform formula $E_m= 2\lambda \cos(2\pi \alpha m)$ for any $\lambda$, the behavior of eigenstates are quite different for $|\lambda|<1$ or $|\lambda|>1$.

Now we consider the case under the GBC. In the region of $\lambda < 1$, for example, $\lambda=0.5$, all eigenstates are localized at the right edge, as displayed in Fig.\ref{uniAA}(b1).
From the finite-size analysis, we find that the localization length is proportional to the system size $L$, as shown in the inset of Fig.\ref{uniAA}(b1).
This can be understood from the self-consistent equation (\ref{bdE}). Using the definition of the LE, Eq.(\ref{bdE}) can be alternatively represented as $\eta e^{- L\gamma} =1$, which gives
\begin{equation}
    \gamma=\frac{\ln(\eta)}{L}.
\end{equation}
It indicates that $\gamma$ is inversely proportional to the system size $L$. In other words, the localization length is proportional to the system size, and the eigenstates are SFL.

In this region, the eigenvalues are complex as shown in  Fig.\ref{uniAA}(c1). In the large size limit, the complex eigenvalues can be analytically determined by $ \gamma= \ln |\frac{E+\sqrt{E^2-(2\lambda)^2}}{2}|$, which gives rise to
\begin{equation}
\frac{E_r^2}{[1+(2\lambda)^2/(4 e^{2\gamma})]^2}+\frac{E_i^2}{[1-(2\lambda)^2/(4 e^{2\gamma})]^2}=e^{2\gamma}
\label{EvW1}
\end{equation}
as long as $\ln(|\lambda|)< \frac{\ln(\eta)}{L}$, where $e^{\gamma}=\eta^{\frac{1}{L}}$.
Equatio(\ref{EvW1}) suggests that the complex spectrum forms an ellipse. The numerical results presented in Fig.\ref{uniAA}(c1) verify the spectrum distribution predicted by Eq.(\ref{EvW1}).
The effect of the boundary parameter $\eta$ to eigenstates is shown in Fig.\ref{uniAA}(d1).
By varying the strength of $\eta$, we observe the transition from "left SFL" to extended states and "right SFL".
The sign of $\gamma$ is determined by $\ln(\eta)$, and $\eta=1$ is the transition point between the left SFL and the right SFL.

In the region of $\lambda>1$,  for example $\lambda=1.5$, the eigenstates are localized states which decay exponentially in a unidirectional way, as shown in Fig.\ref{uniAA}(b2). For a given $\lambda$, all eigenstates have the same LE $\gamma=\ln(|\lambda|)$. Due to the existence of boundary terms, the distribution of the eigenstate $u_i$ in the region $i< m$ is no longer 0 and keeps exponential decay in a unidirectional way. While the eigenstate distributes continuously at the boundary for $\eta=1$, the eigenstates exhibit a sharp decrease and increase at the boundary for $\eta=10^{-4}$ and  $\eta=10^{4}$, respectively, as shown in  Fig.\ref{uniAA}(d2) .

In Fig.\ref{uniAA}(c1),  we present the energy spectrum for $\lambda = 1.5$ and various $\eta$. All eigenvalues are distributed along the real axis and are very close to each other. 
Although the eigenvalues of localized states are not sensitive to the boundary condition, the eigenvalues are no longer given by $2\lambda \cos(2\pi \alpha m)$  in the presence of the boundary term ($\eta \neq 0$). Deviation from the on-site potential can be estimated by Eq.(\ref{bdE}), which can be simplified as:
\begin{align}
 \Delta E_m =E_m-2\lambda \cos(2\pi \alpha m) \approx e^{\ln(\eta) - \ln(|\lambda|) (L-1)}.  \label{DeltaE}
\end{align}
By using this equation, we can estimate the dependence of  $\Delta E_m$ on $\eta$ and the scaling behavior of $\Delta E_m$. In Fig.\ref{uniAA}(c2), we show $\overline{\Delta E}$ versus $\eta$ for different $L$, where $\overline{\Delta E}=\frac{1}{L}\sum_{m=1} ^L \Delta E_m $. The scaling behavior obtained from numerical results are consistent with our analytical analysis. Equation(\ref{DeltaE}) holds true as long as $\ln(|\lambda|)> \frac{\ln(\eta)}{L}$.
The details about the analytical discussion can be found in the Supplementary Material \cite{SM}.

Our results unveil that SFL states manifest when $|\lambda|<\eta^{1/L}$, whereas the system enters the AL region when  $|\lambda|>\eta^{1/L}$.
The SFL-localization transition accompanies a transition from a complex spectrum to a real spectrum. We note that the transition point $|\lambda_c|=\eta^{1/L}$ is size dependent.  For a given $\eta$, when $L \rightarrow \infty$, $\eta^{1/L} \rightarrow 1$.

\begin{figure}[h]
    \centering
    \includegraphics[width=\linewidth, ]{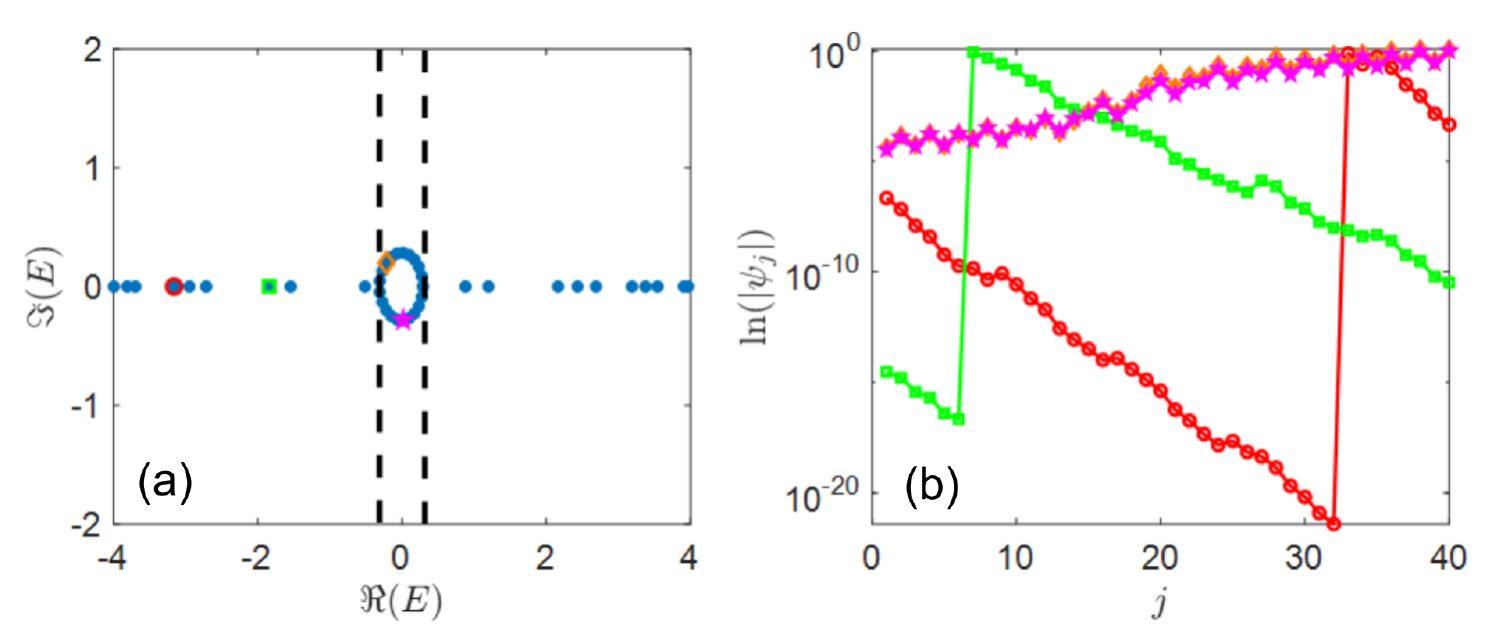}
    \caption{(a) Eigenvalues for the unidirectional mosaic quasiperiodic model with the GBC. The dashed lines indicate the predicted SFL-AL edges.  (b) Distributions of typical eigenstates for the unidirectional mosaic quasiperiodic model. The curves of different colors and markers correspond to the eigenvalues of the corresponding colors and markers in panel (a). Here we choose $\lambda=2$, $\eta=10^{-4}$, $\kappa=2$ and $L=40$. }
    \label{uni_m_AA}
\end{figure}

{\color{blue}\textit{SFL-AL edges in the mosaic model.}}

Next we consider a  unidirectional mosaic quasiperiodic model with the Hamiltonian given by:
\begin{equation}
H_{\kappa}=\sum_{j=1}^{\kappa L-1} t \left\vert
j+1\right\rangle \langle j|+\eta \left\vert 1\right\rangle \langle \kappa L|+ \sum_{j=1}^{\kappa L}   \lambda_j |j\rangle \langle j|,
\end{equation}
where $\lambda_j= 2 \lambda \cos(2 \pi \alpha j), \quad \text{if} \mod(j,\kappa)=0$ and $\lambda_j=0$  otherwise.
By using iteration relations, we can find that the distribution of eigenstates at different sites fulfills:
$$u_{j+\kappa-1}=\frac{1}{E^{\kappa-1}[E-2\lambda \cos(2 \pi \alpha j)]} u_{j-1}.$$
From the bulk equation, we can get the expression of the LE:
\[
\gamma =
\begin{cases}
\ln(|\lambda|)+ (\kappa -1) \ln(|E|),\quad  \text{If}  \quad |E| \leq  2\lambda,  \Im(E)=0\\
\ln |\frac{E+\sqrt{E^2-(2\lambda)^2}}{2}|+ (\kappa -1) \ln(|E|),\quad  \text{others}.
\end{cases}
\]

Considering the boundary condition, we can derive the SFL-AL edge given by:

\begin{align}
\Re(E)=\pm \frac{\eta^{\kappa/L(\kappa-1)}}{|\lambda|^{1/(\kappa-1)}} .
\end{align}

All eigenstates with $|\Re(E)|< \frac{\eta^{\kappa/L(\kappa-1)}}{|\lambda|^{1/(\kappa-1)}}$ are SFL states and eigenstates with $|\Re(E)|>\frac{\eta^{\kappa/L(\kappa-1)}}{|\lambda|^{1/(\kappa-1)}}$ are localized\cite{SM}. Here the SFL-AL edge is size dependent.
Different from the Hermitian case, the multiple mobility edges emerging in the Hermitian mosaic model with $\kappa \geq 3$ \cite{WangYC} disappear in the unidirectional model.
In Fig.\ref{uni_m_AA}(a), we present the results from the exact diagonalization and the exact SFL-AL edge predicted by our method. In Fig.\ref{uni_m_AA}(b), we show some typical eigenstates which exhibit SFL or AL behavior. For $\eta=1$, the mobility edge is given by $\Re(E)=\pm |\lambda|^{-1/(\kappa-1)}$, which separates the extended states from the localized states.

{\color{blue}\textit{Summary and discussion.}} We studied the effect of quasiperiodic potentials in a unidirectional hopping lattice with GBC. By calculating the LE analytically,
we explored the interplay between SFL and AL and identified the transition point from SFL to AL is given by $\eta^{1/L}$.  In different phase regions, two types of eigenstates satisfy the boundary condition through different mechanisms. The SFL eigenstates are characterized by the localization length $\xi=L/|\ln(\eta)|$  and acquire complex eigenvalues which distribute on an ellipse. In contrast,
the localized eigenstates  are characterized by the localization length $\xi=\ln|\lambda|$ and have real eigenvalues which align closely with the on-site potentials. 
Further generalizing our framework, we designed a model with accurate SFL-AL edges via the mosaic modulation of quasiperiodic potentials. Our theoretical scheme can be directly extended to study other  unidirectional quasiperiodic systems.

Our analytical results establish exact relationships between parameters ($\eta$ and $\lambda$) and behaviors of the system. Our study unveils the important role of the boundary term in the formation of SFL and the estimation of spectrum properties of localized states, deepening our understanding of the underlying mechanisms of transition between SFL and AL.
We can also design electric circuits to simulate our models \cite{SM}, which helps us generalize the theory to practical applications.


\begin{acknowledgments}
We thank Qi Zhou and Mingchen Zheng for useful discussions. This work is supported by National Key Research and Development Program of China (Grant No. 2023YFA1406704)
and the NSFC under Grants  No.12474287  and No. T2121001. 
\end{acknowledgments}

\end{document}